\documentclass[fleqn,usenatbib]{mnras}

\usepackage{newtxtext,newtxmath}

\usepackage[T1]{fontenc}

\DeclareRobustCommand{\VAN}[3]{#2}
\let\VANthebibliography\thebibliography
\def\thebibliography{\DeclareRobustCommand{\VAN}[3]{##3}\VANthebibliography}

\usepackage{comment}
\usepackage{graphicx}	% Including figure files
\usepackage{amsmath}	% Advanced maths 
\title[]{
Explaining the "too massive"  high-redshift galaxies in JWST data: \\12a numerical study of three effects and a simple relation
}

\author[J. J. Ziegler et al.]{
Joshua J. Ziegler,$^{1}$\thanks{E-mail: jjziegler@utexas.edu}
Katherine Freese$^{1,2,3}$, Jonathan Lozano$^{1}$, and
Gabriele Montefalcone$^{1}$ 
\\
$^{1}$Texas Center for Cosmology and Astroparticle Physics, Weinberg Institute for Theoretical Physics, \\ \,\,\,Department of Physics, The University of Texas at Austin, Austin, TX 78712, USA\\
$^{2}$The Oskar Klein Centre, Department of Physics, Stockholm University, AlbaNova, SE-10691 Stockholm, Sweden\\
$^{3}$Nordic Institute for Theoretical Physics (NORDITA), 106 91 Stockholm, Sweden
}

\date{Accepted XXX. Received YYY; in original form ZZZ \\
\\
Preprint Numbers: NORDITA-2025-028, UT-WI-18-2025}

\pubyear{\the\year{}}

\begin{document}
\label{firstpage}
\pagerange{\pageref{firstpage}--\pageref{lastpage}}
\maketitle

% Abstract of the paper
\begin{abstract}
The James Webb Space Telescope has discovered high luminosity galaxies that appear to be ``too many'' and ``too massive'' compared to predictions of the Standard 
$\Lambda$CDM cosmology, suggesting that star formation in the early universe is more rapid than previously anticipated. 
In this paper we examine in detail the following three effects which can instead provide alternative explanations for these observations: 
(1) a ``top heavy'' initial mass function (IMF) for the stars (high mass stars produce far more light than low mass stars),
(2) a variety of star formation histories (constant, exponentially decreasing, and peaked star formation rates), and (3) a variety of initial metallicities.
Due to any of these three effects, galaxies of a given luminosity in JWST may be interpreted as having a larger stellar mass than they actually do.
Our results are obtained using the \textsc{P\'egase} stellar population code, and are presented as the ratio of the modified star formation efficiency  relative to the fiducial one (which uses
a Salpeter IMF and constant star formation rate). 
As an example, 
if the high-mass end of the IMF goes as $M^{-1.35}$, the star formation efficiency and inferred stellar galactic mass could be lower by a factor of $\sim 10$ than in the fiducial case.
Our examination (keeping the star formation rate constant) of a top-heavy IMF with slope $\alpha$ (where $\alpha = 2.35$ for Salpeter IMF) leads to
 a simple relation that is a good approximation to the numerical results, $\epsilon(\alpha) \approx \epsilon_{\rm fid}e^{2.66(\alpha -2.35)}$,
 corresponding to an inferred stellar galactic mass $M_{\rm lum}(\alpha) \approx M_{\rm fid}\, e^{2.66(\alpha - 2.35)}$, given an observed galactic luminosity.
Since there are more low mass galaxies than high mass galaxies, these effects may result in a large number of seemingly overly massive galaxies compared to the expectations.
Thus, the effects studied in this paper may explain both puzzling observations regarding high luminosity galaxies in JWST:
the apparently overly massive galaxies as well as the profusion of apparently high mass galaxies.
\end{abstract}

\begin{keywords}
galaxies: luminosity function, mass function -- galaxies: high redshift -- galaxies: star formation -- surveys
\end{keywords}

\section{Introduction}

The James Webb Space Telescope is observing galaxies at higher redshifts than ever before. 
These observations have led to 
surprising conclusions: galaxies are more abundant and several galaxies more massive than previously predicted based on the standard LCDM cosmology~\citep{Boylan_Kolchin_2023, Labbe2023}. 
As a result, these observations were initially thought to ``break'' the standard LCDM model of cosmology.
In particular, initial analysis of the brightest early galaxies seemed to require nearly all of the baryons in a halo to collapse very rapidly into stars,
with an efficiency of gas conversion to stars of $\epsilon \sim 100\%$~\citep{Boylan_Kolchin_2023,lu2024, Shuntov_2025}.
The (time-averaged) star formation efficiency $\epsilon$ (SFE) is related to the stellar mass in the galaxy $M_{\rm lum}$ via
\begin{equation}
M_{\rm lum} = \epsilon f_b M_{\rm halo} = \epsilon M_b \label{eq:1}
\end{equation}
where $f_b$ is the cosmic baryon fraction, $M_{\rm halo}$ is the total galactic mass, and $M_b$ is the total baryonic mass in the galaxy~\citep{myers1986}.
However, more recent studies of the most extreme galaxies~\citep{chworowsky2023, woodrum2023} estimate a lower efficiency rate of $\epsilon \sim  30-40\%.$ Although such an efficiency resolves the immediate contradiction to LCDM, it is
still significantly in excess of star formation efficiencies observed in nearby galaxies
($\epsilon \sim 1-10\%$), and remains challenging to explain astrophysically. 

In light of this challenge, many explanations for the unexpectedly large SFE at high redshifts have been proposed. While these proposals span a broad range of phenomena, we can loosely divide them into three categories. (1) Star formation in high-redshift environments may occur through mechanisms that are inaccessible to low-redshift galaxies, including 
rapid cycles of star formation bursts and quiescence~\citep{Pallottini2023,sun2023, Casey2024} and dynamic funneling of cold gas into a dense star forming region~\citep{Guo2023, mckinney2024}. (2)  The total mass of baryons associated with high-redshift galaxies could be underestimated. For example, these galaxies could (through random variability) have a higher baryon to dark matter ratio and would therefore be brighter and more easily visible than galaxies at comparable redshifts~\citep{shen2023}. 
(3)  In order to estimate the properties of high-redshift galaxies, it is necessary to match observed spectra or luminosity \footnote{A resolved spectrum of a galaxy can be produce the most reliable estimates for some properties of a galaxy (including mass and star formation efficiency), but for distant galaxies spectra cannot always be resolved. In these cases photometric measurements which integrate the luminosity in a wider frequency band can be used to approximate the spectral energy distribution of a galaxy, but with higher uncertainties.} to a library of theoretical stellar populations~\citep{Pacifici_2023}. It is possible that the libraries used in parameter estimation are inaccurate, for example the assumed initial mass function or dust absorption models may not be correct~\citep{Steinhardt_2023}. 
In this paper, we study the latter of these possibilities and explore how modifying assumptions made when producing the fitting libraries affects the observed star formation efficiency.

In particular, one of the most promising quantities
is the Initial Mass Function (IMF), the number of stars as a function of mass of the star at the time of their formation,
\begin{equation}
    dN_\star/dM_\star \propto M_\star^{- \alpha}. \label{eq:2}
\end{equation}
For stars in the local universe, the IMF has been observed to follow a common power law (for stars with mass $M_\star \gtrsim 0.5 M_\odot$) known as the Salpeter mass function, which has
$\alpha_{\rm fid} = 2.35$~\citep{Salpeter1955}. 
However, it is expected that high-redshift galaxies will exhibit a more top-heavy IMF, in which case $\alpha < \alpha_{\rm fid}$.

While our primary focus is on the role of the IMF, in this paper we quantitively examine the role of each of the following three quantities on the stellar mass of a galaxy that can be extracted from its observed luminosity
via their effects on the early star formation efficiency (SFE): \newline
1) IMF power law $\alpha$, in the range $\alpha\in[1.35,\, 2.35]$, where we use Salpeter's 2.35 as our fiducial value. \newline
2) initial metallicity $Z_{\rm ini}$, in the range $Z_{\rm ini}\in[10^{-3},\, 1]$, where we use $10^{-3}$ as our fiducial value. \newline
3) Star Formation History (SFH), i.e. the time variation of the Star Formation Rate, ${\rm SFR}(t)$.
We consider three SFHs: constant, exponential, and peaked, as defined further below, and use the constant SFH as our fiducial behavior.

We use the population synthesis code \textsc{P\'egase}~\citep{Fioc_2019} to model spectral energy distributions (SEDs) of galaxies
with the above choices for the three sets of parameters.  
The outputs of the simulations that are of interest in this work are, as a function of the time parameter, i.e. the age of the galaxy, \newline
1) the stellar mass of the galaxy $M_{\rm lum} (t)$ \newline
2) the metallicity $Z(t)$ \newline
3) the luminosity of the galaxy as a function of time and frequency $L(t, f)$ where the frequencies correspond to filters in JWST.

An important component of our paper is assessing how the star formation efficiency inferred from observations of a high-redshift galaxy is altered if different IMFs are assumed.
As expected, a ``top-heavy'' IMF improves the match between the observed high luminosity objects in JWST and the predictions of LCDM without requiring exotic star formation mechanisms.
As $\alpha$ decreases in the figure, so does $\epsilon$ and thus $M_{\rm lum}$ from Eq.~\eqref{eq:1}.
To a rough approximation, the luminosity of an individual star scales roughly as  $L_\star \propto M_\star^{3.5}$. 
In turn, high mass stars produce far more light per unit mass than low mass stars, so that a lower total stellar galactic mass is required to 
produce the observed luminosities of high z galaxies.
As a result, the luminosity emitted by a galaxy will depend both on the total mass of that galaxy and how the mass is distributed amongst different stars.
Because more of the stars within a galaxy with a top-heavy IMF are massive, galaxies with a more top heavy IMF ($\alpha < \alpha_{\rm fid}$) produce the same light with less overall total stellar mass (all other factors held equal). In turn, when determining the mass of a galaxy from observed luminosity data, assuming a more top-heavy IMF in the analysis will result in lower estimates for the galaxy's mass.

Many authors have explored the effect of top heavy IMFs on the SFE in early JWST galaxies,
including
\citet{trinca2024, Trinca2022, Cueto_2024, menon2024, woodrum2023, lapi2024}.

One of our main results is the following simple relationship between the IMF parameter $\alpha$ and the SFE,
\begin{equation}
    \epsilon = \epsilon_{\rm fid} e^{2.66(\alpha- 2.35)}
\end{equation}
where $\epsilon_{\rm fid} $ is the SFE corresponding to the  Salpeter IMF.
This relationship offers a shortcut to 
more complex analyses involving SED fitting.
It can also simplify the analysis of the systematic uncertainty in the SFE due to the IMF, so that this factor is not overlooked. 

In section~\ref{sec:galev}, we discuss the standard picture of star formation as well as present our parameter choices for the three effects studied in this paper.
In section~\ref{sec:sfe}, we discuss the star formation efficiency in these galaxies and how it can be used as a metric for the impact of varying the IMF. In section~\ref{sec:methods}, we discuss in more detail how we use \textsc{P\'egase} to model the SEDs of galaxies with different IMFs. In section~\ref{sec:results}, we present the results from our analysis including the relationship we compute between the IMF and star formation efficiency. Finally, we conclude in section~\ref{sec:conc}.

\section{Theoretical Considerations and the Parameters Explored in this Paper}
\label{sec:galev}

First, we discuss factors that impact the formation of stars in a galaxy and their mass distribution. We focus primarily on the initial mass function, the star formation history, and the initial metallicity of the gas in the galaxy. Galactic emission spectra depend sensitively on each of these factors, and in turn galactic properties inferred from spectra are also sensitive to these factors. 

In the standard LCDM model of cosmology, galaxy formation begins with the growth of overdensities from the early universe.  On large scales,
the linear growth of dark matter halos can be quantified using the Press-Schecter formalism~\citep{1974press-schechter}, which provides a method for counting the number of halos below a given mass at a given redshift. For a redshift $z$, at some mass $m_{c}^{\rm max}(z)$ the number of halos must be less than one in the observable universe. This mass $m_{c}^{\rm max}(z)$ is thus the maximal halo mass at the redshift $z$.  
In turn, the expected maximal mass of baryons bound within any dark matter halo at redshift $z$ is $m_{b}^{\rm max}(z) = (\Omega_b/\Omega_c) m_{c}^{\rm max}(z)$, where $\Omega_b$ and $\Omega_c$ are the baryon mass fraction and dark matter mass fraction~\citep{Boylan_Kolchin_2023}. 
As galaxies form, baryons (in gas and dust) coalesce into stars, which are expected to be the largest source of ultraviolet radiation that JWST is sensitive to in many high redshift galaxies.
However, with standard assumptions about star formation and stellar evolution, all of the baryonic material in the largest dark matter halos would have to be rapidly converted to stars to produce the high luminosity seen by JWST in high redshift galaxies.
Equivalently, inferring the mass of the high-redshift galaxies observed by JWST using standard assumptions about star formation and evolution can lead to inferred galaxy masses that exceed the total baryonic mass in the highest-mass haloes at the relevant redshift.

\subsection{Initial Mass Function}

The conclusion outlined above, that several high-redshift galaxies observed by JWST have masses greater than would be expected from LCDM, depends heavily upon the assumptions made about the stellar content of these galaxies. To a rough approximation, the luminosity of an individual star is $L_\star \propto M_\star^{3.5}$, so a single massive star will produce more light than a collection of lower-mass stars whose total mass is the same as the single star. As a result, a galaxy composed of predominantly massive stars will have a greater overall luminosity than a galaxy with the same mass but composed of predominantly lower-mass stars. This distribution of stellar masses within a galaxy is related to the initial mass function of the galaxy. For the purposes of this project, we consider initial mass functions whose high-mass behavior can be characterized as a power law with a power $\alpha$, as defined in Eq.~\eqref{eq:2}.

The reason we focus on power-law IMFs can be traced to the manner in which clusters of stars form from within giant molecular clouds~\citep{offner2014}. Overdense regions within molecular clouds collapse to form clumps. As these clumps collapse further, their temperature generally increases. In order for these clumps to collapse enough for nuclear fusion to begin, this heat built up during the gravitational collapse must be expelled. However, the cooling efficiency of a clump is sensitive to, for example, the clump's metallicity. Stars formed in relatively metal-rich clouds ($Z_{\rm ini}\gtrsim0.01 Z_\odot$), including young Population I stars seen in the Milky Way and nearby galaxies, can expel heat efficiently through metal-line cooling methods~\citep{peters2014}, allowing fragmentation to smaller objects~\citep{offner2014}. However, because the metals in these clouds are produced by earlier generations of stars, the first star forming regions would have effectively no metals, consisting only of hydrogen and helium from the big bang. For clumps in these early star forming regions, cooling would have occurred by inefficient molecular hydrogen reactions. By studying nearby star formation, we can identify an empirical IMF for stars that form using efficient metal-line cooling. This IMF, first observed by Salpeter, is of the form $dN_\star/dM_\star \sim M_\star^{-2.35}$ i.e. $\alpha = 2.35$~\citep{Salpeter1955}. The analogous study of star formation in extremely metal-poor environments has not yet been possible, so the exact nature of the IMF in these environments remains uncertain. However, if we assume the process of star formation in these metal-poor environments is analogous to that in more metal-rich environments, a reasonable expectation is that the metal-poor IMF would also approximate a power law, but with $\alpha<2.35$ as a result of the more inefficint cooling. 

Because metallicity tends to increase with time, the average IMF will evolve with redshift. Lower metallicity at higher redshift (on average) implies that the typical IMF of high redshift galaxies is likely to be more top-heavy than the Salpeter IMF. In this paper, for stars heavier than $0.5 M_\odot$, we take $\alpha$ to vary over the range $\alpha\in[1.35,\, 2.35]$, to explore a range of top-heavy power-laws that could reasonably be expected in high-redshift galaxies.

\subsection{ Star Formation History}
\label{sec:SFH}

In addition to the IMF, different star formation histories also have a significant impact on the emission spectra of galaxies. Because the  star formation rate depends on, for example, the rate at which gas can be accreted onto galaxies from the circumstellar medium, it is possible for the star formation rate to change over time. This evolution of the star formation rate is the star formation history. For the purposes of this paper, we will consider three common classes of SFH: constant, exponential, and peaked. 
 \begin{itemize}
    \item \emph{Constant}: Here, the star formation rate is taken to be constant over the entire history of the galaxy, normalized such that the total mass of the galaxy is converted to stars over the course of 20,000 Myr. Specifically, the star formation rate is ${\rm SFR}_{\rm const} = 1.395\times10^{6}~\mathrm{M_\odot \, Myr^{-1}}.$ 
    \item \emph{Exponential}: The exponential SFH that we use assumes an instantaneous onset to star formation followed by a star formation rate that decreases exponentially with time as ${\rm SFR}_{\rm exp}\propto \exp(-t/\tau)$ with time constant $\tau$ taken to be either 10 Myr or 100 Myr. In either case, the normalization is such that the total mass of the system is converted to stars over the same 20,000 Myr lifetime.
    \item \emph{Peaked}: The SFR is taken to initially rise linearly with time (rather than having an instaneous onset as above), followed by the same exponential decrease at late times:
     ${\rm SFR}_{\rm peak} \propto t \exp(-t/\tau)$, where we again choose $\tau$ to be either 10 or 100 Myr.
     This case is particularly useful in comparison with the \emph{exponential} case to model the effect that an initial increase in the star formation rate would have on observables.  As in the \emph{exponential} case, we normalize the star formation rate under the condition that the entire mass of the system is converted to stars over the course of 20,000 Myr. 
\end{itemize}
Star formation histories in high-redshift galaxies remain hotly debated. Locally, star formation histories tend to approach the peaked case~\citep{Speagle2014}, but the star formation histories of early galaxies cannot necessarily be assumed to follow this same pattern. For example, if the rate at which gas falls into a galaxy from the circumgalactic medium is roughly constant in early galaxies, the star formation rate would also be approximately constant. Alternatively, if star formation tends to form in brief bursts followed by quiescent periods, the star formation during the longer quiescent period could approximate an exponential decline, though if the exponential timescale is relatively long, it may be difficult to distinguish an exponentially decaying star formation rate from a constant star formation rate in high-redshift galaxies.

Finally, in addition to the star formation history, stellar evolution effects also influence the emission spectra of galaxies over time. In general, more massive stars have shorter lifetimes than their less massive counterparts~\citep{kippenhahn}. This means that if star formation in a given star-forming region extends over a period of time, the distribution of stellar masses within that star-forming region will not follow the IMF that those stars form with. In fact, the stars in a star-forming region will typically contain more low-mass stars than naively predicted by the IMF. Furthermore, as more massive stars die, they tend to enrich their surroundings with metals, which can affect the nature of the IMF for subsequent generations of stars.

\subsection{Metallicity}
As a galaxy ages, early generations of stars enrich the gas in the galaxy with metals, which then influences star formation and evolution of later generations of stars.
To isolate  the role of metallicity, in addition to the fiducial case with initial metallicity $10^{-3}~Z_\odot$, we consider galaxies with constant star formation histories and unphysically large initial metallicities $Z_{\rm ini}$ of $0.1~Z_\odot$ and $1~Z_\odot$,
and will show the dependence of the results on these choices.  

Given a set of assumptions about the IMF, the star formation history, and the initial metallicity, it is possible to model the population statistics of stars within a given star-forming region at any given time after stars have begun forming. From such a population, it is possible to predict the stellar emission spectrum. If we then make assumptions about the types of absorption and re-emission that this stellar emission spectrum undergoes (e.g. by dust), we can predict the spectral energy distribution (SED) for that population. However, when observations of galaxies are made, the inverse process of this forward modeling must be performed. This SED fitting process unfortunately includes numerous parameters that are partially or completely degenerate, giving rise to potential systematic uncertainties in the estimation of parameters fit using this technique. Instead, in this paper we take a simpler approach as we  describe below.

\section{Obtaining Star Formation Efficiency from Observed Luminosity}
\label{sec:sfe}
When observing galaxies, especially at high redshift, it is effectively impossible to directly observe star formation efficiency or any of the parameters that influence star formation. As such, we are limited to spectra, and in some cases only photometric data, or even only an overall luminosity. In light of this, we outline how we can relate the luminosity of a galaxy to its star formation efficiency. In the process, we will highlight generally how altering star formation parameters from fiducial values can impact the relationship between luminosity and star formation efficiency.

First, we discuss the conventional approach to parameter estimation based on emission spectra. Large libraries of spectral energy distributions (SEDs) are constructed by simulating star forming regions with a variety of different parameters. Then observations of galaxies are fit to these libraries of SEDs, with the best fit providing an estimate for the mass of stars in the galaxy, and a corresponding star formation efficiency. However, many of the parameters varied in the construction of SED libraries produce similar effects on the SEDs, and many parameters are sparsely sampled. For example, in many libraries, the default choice is to only consider a Salpeter (or Salpeter-like) IMF, in part because adding additional IMFs to the library significantly increases the complexity of the fitting without improving the fit, potentially because of strong degeneracies between the IMF and other fitting parameters. A constant star formation history is commonly used for similar reasons.

We turn now, to the simpler approach we take. Given an observed luminosity of a galaxy, we will use estimates of its mass-to-light ratio ${(M/L)_{\rm gal}}$  to obtain a value for its stellar mass $M_{\rm lum}$.  Knowing $M_{\rm lum}$ is then equivalent to knowing the SFE $\epsilon$ via Eq.~\ref{eq:1}. To obtain  
the mass-to-light ratio needed for this process, we initially outline in this section a simple analytic estimate for the initial $(M/L)_{{\rm gal}, \, {\rm ini}}$  for illustrative purposes;
then in the next section, we will discuss how we use the \textsc{P\'egase} code to numerically model galactic spectral emission distributions and corresponding mass-to-light ratios.  The mass-to-light ratio for a galaxy depends on a number of parameters, including the IMF, SFH, and metallicity that we explore.

First, we look at the bolometric luminosity emitted by a galaxy. 
Assuming that all light from a galaxy is produced initially by stars, and then potentially absorbed and re-emitted by gas and dust, the bolometric luminosity $L_{\rm bol}$ is directly correllated with the stellar mass:
\begin{equation}
\label{eq:mass-to-light}
    M_{\rm lum} = {(M/L)_{\rm gal}} L_{\rm bol}.
\end{equation}

The mass-to-light ratio $M/L$  is determined both by the IMF and the age of the galaxy. When the first stars in a galaxy are formed, the galactic mass to light ratio can be calculated as $(M/L)_{{\rm gal},{\rm ini}} = \int {(dN_\star \over dM_\star}) (M/L)_\star dM_\star$, where $(M/L)_\star$ is the mass-to-light ratio of a star of mass $M_\star$. We can approximate  $(M/L)_\star \sim M_\star^{2.5}$,  so that initially (subscript ``${\rm ini}$'')
\begin{equation}
    (M/L)_{{\rm gal},{\rm ini}} \approx \frac{1}{\alpha + 3.5} \left( M_{\star, \, \rm upper}^{\alpha+3.5} - M_{\star, \, \rm lower}^{\alpha+3.5} \right),
\end{equation}
where $M_{\star, \, \rm upper}$ and $M_{\star, \, \rm lower}$ are the upper and lower mass bounds considered for the IMF, respectively. 
While useful to give a rough estimate for the initial mass-to-light ratio of a galaxy, this expression quickly becomes insufficient for our purposes as stellar evolution drives the stellar population away from the initial mass function.
Shortly after this initial star formation, the stars with the highest masses begin to explode in supernovae, and consequently no longer contribute to the mass-to-light ratio. Because more massive stars have shorter lifetimes than lower mass stars, at any time after the lifetime of the most massive stars, the stellar distribution in the galaxy will no longer be described by the  IMF. The exact mass distribution of the stars in a galaxy will depend on the precise star formation history, particularly how the rate of star formation compares with the rate of deaths of stars of different masses. Furthermore, as stars die and expel metals into the interstellar medium, the stars that form at different times will generally have different chemical composition. Ultimately, this means that the approximation above will no longer be valid. Instead, as described in the next section,
we need to follow the population statistics numerically to predict the mass-to-light ratio of the galaxy at any time later than the initial star formation.

Finally, observationally, we rarely if ever measure the bolometric luminosity. Instead, we measure the luminosity through a finite band, and from this infer the bolometric luminosity. We can describe this as 
\begin{equation}
\label{eq:ffilter}
    L_{\rm filter} = f_{\rm filter}(\nu,z) L_{\rm bol},
\end{equation}
where $f_{\rm filter}$ is the fraction of the galaxy luminosity that is emitted in the filter band. Note that in this paper, rather than considering individual JWST filters, we will
add up all the light from the entire frequency range observed by JWST.  Hence, in the rest of this paper,  $L_{\rm filter}$ will include the total luminosity observed in JWST.
The filter fraction  depends only on the redshift of the galaxy ($z$) and the frequency range of the filter ($\nu$). With this in mind, we define
\begin{equation}
\beta \equiv {M_{\rm lum} \over L_{\rm filter}} = \frac{(M/L)_{\rm gal}}{f_{\rm filter}} \label{eq:7}
\end{equation}
where the second equality follows from Eq.~\eqref{eq:mass-to-light} and~\eqref{eq:ffilter}.
Thus, parameter $\beta$ contains both the mass-to-light ratio effects and filter effects. Using the definition of $\beta$ together with Eq.~\eqref{eq:1},  the SFE can then be identified as 
\begin{equation}
    \epsilon =\frac{\beta}{M_b}  L_{\rm filter} = \frac{(M/L)_{\rm gal}L_{\rm filter}}{f_{\rm filter} M_b}
\end{equation}

In this paper, our goal is to examine the effects of varying three quantities, the IMF, the star formation history, and the initial metallicity,
on this process. We will examine how these variations modify
the value of galactic stellar mass $M_\star$, or equivalently $\epsilon$ inferred from
the measured luminosity of a galaxy, i.e. the luminosity in the relevant filter range of the observation. 
Each combination of parameters
gives rise to a different value for the SFE $\epsilon$ and for $M_{\rm lum}$.
We will present our results in terms of ratios of $\epsilon/\epsilon_{\rm fid}$ where the denominator corresponds to the fiducial results obtained
as described above with the SED fitting with the standard assumptions about the three quantities of interest.
Regarding the IMF, we take the Salpeter IMF to be the fiducial value in the denominator of $\epsilon/\epsilon_{\rm fid}$, and find the ratio that determines how much smaller (larger) the inferred SFE would be if a more top-heavy (bottom-heavy) IMF were assumed for that galaxy. We do the same procedure for the other parameters we vary as well.

\section{Numerical Approach}
\label{sec:methods}

In order to quantify the effects of varying the star formation parameters described earlier on the SFE and resulting galactic stellar mass $M_\star$,
we use the stellar population synthesis code
\textsc{P\'egase 3} ~\citep{Fioc_2019}. This code allows us to simulate the emission spectra and stellar properties of galaxies as we vary star formation parameters. Our primary interest is, as above, the initial mass function, the star formation history, and the initial metallicity, so we use \textsc{P\'egase} to explore how stellar emission spectra and stellar mass vary as we vary these input parameters.  For fiducial values of these parameters, the SFE and stellar galactic mass $M_{\rm lum}$~(see Eq.~\eqref{eq:1}) are known to 
be uncomfortably large for the standard LCDM model and star formation models, and we will quantify how varying the three parameters changes these results.

We here describe a few features of the code.  \textsc{P\'egase} considers a three-environmment model: a reservoir of cool gas from which stars are formed, a galaxy in which stars form and evolve, and a reservoir of hot gas produced by outflows. The reservoir of cool gas is treated as fully independent from the reservoir of hot gas. 
Throughout our analysis, we consider a spheroidal galaxy contained within a halo whose total baryonic mass content is $2.79 \times10^{11}~\mathrm{M_\odot}$, the default total baryonic mass for spheroidal galaxies in \textsc{P\'egase 3}. However, the analysis we perform can be extended to different masses through simple rescaling, as the total mass normalizes both the mass in stars ($M_{\rm lum}$) and the overall luminosity $L_{\rm bol}$.

We model the evolution of the stars within a galaxy as a series  of single stellar populations (SSPs), each containing stars formed at approximately the same time and with the same initial metallicity, with a mass distribution determined at that time by the IMF. In principle, the IMF can vary over time, and different SSPs would reflect this, but we consider only cases where the IMF is constant over the period of the galaxy's history that we are considering.
However, the mass of stars produced within each SSP also depends on the star formation rate, which we allow to vary over the history of the galaxy. 
As described previously in Section \ref{sec:SFH}, 
we consider three models for the evolution of this star formation rate: constant, exponential, and peaked.
For each SSP that is added to the galaxy, we consider stellar evolutionary tracks based on those computed by the Padova group in the 1990s for stars with initial mass $0.1~\mathrm{M_\odot}$ to $120~\mathrm{M_\odot}$ and described in more detail in section 2.2.2 of \citet{Fioc_2019}. These allow us to identify the spectra of each SSP as a function of time following its formation. The total spectrum from stars within a galaxy at any given time $t$ is then the sum of the spectra from each SSP at the corresponding age: $B(\nu, t) = \sum B_{j}(\nu, t-t_{{\rm ini,j}}),$ where $B_j(\nu, t)$ is the spectrum from SSP $j$ evaluated at time $t$. Furthermore, these evolutionary tracks allow us to determine when stars of different masses will die (either as supernovae or forming white dwarfs). While we consider different initial metallicities of the gas in a galaxy, in all cases we assume that the metallicity of the gas from which new stars form increases as earlier generations of stars enrich the interstellar medium, and the evolutionary tracks and stellar yields (described in section 2.2.3 of \citet{Fioc_2019}) provide a way to follow this enrichment.

Once we have the total emission spectra from stars within the galaxy, we consider how this light is processed through dust. To model dust extinction and emission, we use the ``\textsc{Bare\_GR\_S}"
model of \citet{Zubko_2004}. The dust is heated both by stellar emission and by the cosmic background radiation. 
For specificity, in running the simulation we take the formation redshift of the galaxy to be at $z=13.5$. Under this assumption, the galaxy will be approximately 100 Myr old at redshift $z\sim11$.  Yet, as we now describe, the results we obtain will be essentially the same regardless of the particular reddhift chosen.

Generally, the redshift at which galaxies form only enters into the \textsc{P\'egase} calculations through the cosmic background radiation. Galaxies that form earlier experience a higher temperature cosmic background radiation, which in turn causes the dust in the galaxy to have a higher temperature. Dust temperature affects the absorption properties of that dust, so the increased dust temperature can slightly shift the emission spectrum of a galaxy formed at higher redshifts. However, we are primarily interested in relatively metal-poor (and consequently dust-poor) galaxies, so changing the formation redshift will have minimal effect on the emission spectrum. As such, while the results we present are based on a formation redshift of $z=13.5$, our results should remain unchanged if the galaxy forms at an earlier redshift. For example, a galaxy that forms at redshift $z=19$ and is observed at $z=14$, a galaxy that forms at redshift $z=17$ and is observed at redshift $z=13$, and a galaxy that forms at redshift $z=13.5$ and is observed at redshift $z=11$ will all be approximately 100~Myr old when observed\footnote{An age of approximately 100~Myr is of interest as many previous studies of the observable effects of a top-heavy IMF use this galactic age when calculating galactic emission spectra.}, and therefore will all show similar spectra.

Then, we calculate the quantity $L_{\rm filter}$ by integrating the spectrum over an appropriate frequency, namely the JWST NIRCam filters~\citep{Rieke2008}. The output from \textsc{P\'egase} is the dust-processed spectra of a galaxy at its source redshift. As such, we redshift the filter frequencies to what they would be at a redshift $z=11$ before integrating. In particular, we use a filter with a constant transmission across the wavelength range 83.3--416.7~nm as a simplified approximation to the NIRCam filters. Finally, the mass $M_{\rm lum}$ contained within stars  at any given time can also be found by summing the masses of stars within each SSP that have not yet reached the end of their lifetimes. 
The stellar mass $M_{\rm lum}$ is related to the SFE $\epsilon$ via Eq.~\eqref{eq:1} and $\beta$ via Eq.~\eqref{eq:7}.

\begin{figure*}
    \centering
    \includegraphics[width=\linewidth]{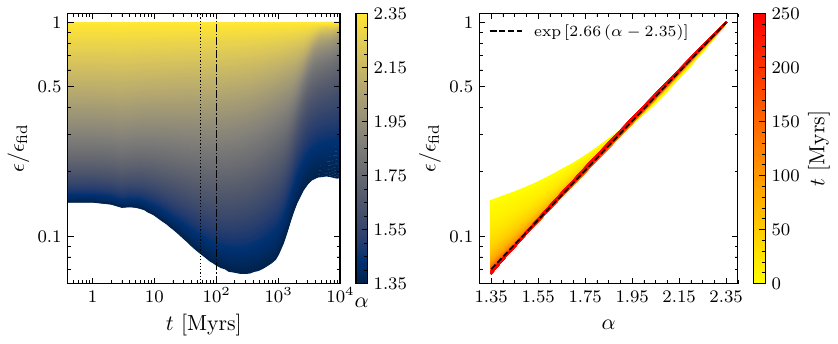}
    \caption{
    {\it Left}: Evolution of the star formation efficiency ratio $\epsilon/\epsilon_{\rm fid}$ (equivalently the inferred mass ratio $M_{\rm lum}/ M_{\rm fid}$) as a function of the age of a galaxy $t$ for various top-heavy IMFs $dN/dM \propto M^{-\alpha}$
 characterized by different high-mass power-law slopes  $\alpha$  ranging from 2.35 (Salpeter-like IMF)  to 1.35. Each IMF is represented by a distinct color as indicated by the colorbar to the right of the panel. All simulations assume an initial metallicity  $Z_{\rm ini}=0.001\,Z_{\odot}$ and a constant star formation rate. The denominator, $\epsilon_{\rm fid}$, corresponds to the fiducial Salpeter-like IMF ($\alpha=2.35$). We note that smaller values of $\epsilon$ correspond to smaller values of stellar galactic mass $M_{\rm lum}$ inferred
 for a given galactic luminosity. The vertical dotted and dash dotted lines mark $t=55\,$Myr and $t=102\,$Myr respectively. The corresponding redshift depends on the formation time of the galaxy, e.g., if  the galaxy formed at $z=13.5$, the vertical lines correspond to redshifts $z\sim 12$ and $z\sim 11$ respectively. {\it Right:} Relation between the ratio $\epsilon/\epsilon_{\rm fid}$ and the high-mass IMF slope $\alpha$ at different galaxy ages, from 10 to 250 Myr shown in a color gradient from yellow  to red. For $t\gtrsim 150\,{\rm Myr}$, the relation stabilizes and is well-described by $\epsilon/\epsilon_{\rm fid}\approx \exp\left[ 2.66(\alpha -2.35)\right]$ (the black dashed line). Together, both panels illustrate that more top-heavy IMFs consistently lead to lower $\epsilon$ ratios, and therefore lower inferred star formation efficiencies and galactic stellar masses for a given observed galaxy luminosity. }
    \label{fig:1}
\end{figure*}

\section{Results}
\label{sec:results}

Our primary goal in this analysis is to evaluate the degree to which the inferred properties of distant galaxies, such as those observed by JWST at redshift $z>10$, depends on the assumptions made about star formation, including the IMF. Physically it is the stellar galactic mass $M_\star$ inferred from the data that is of interest.
This quantity is directly related to the Star Formation Efficiency $\epsilon$ via Eq.~\eqref{eq:1}. 
 We begin with a fiducial case, where we assume the initial metallicity $Z_{\rm ini} = 0.001 Z_\odot$, a constant star formation history, and a Salpeter-like IMF. Specifically as our fiducial case we take an IMF of the form $dN_\star/dM_\star \propto M_\star^{-\alpha}$, where $\alpha$ is fixed to $\alpha = 1.3$ for $M_\star<M_\odot$ and $\alpha = 2.35$ for $M_\star\geq0.5  M_\odot$ in the fiducial case. 

For specificity, we study galaxies that form at $z_{\rm form} = 13.5$. However, galaxies that form at other high redshifts should show similar results.
 The specific values of galactic age shown in our plots would correspond to shifted values of the redshift, but the conclusions would remain the same.
 
 \subsection{Varying the IMF}
 
 We  study the effect that varying the IMF has on the SFE and consequently $M_\star$ deduced from JWST observations of a galaxy.
We  take the slope of the IMF for $M_\star\geq0.5 M_\odot$ to vary within the range $\alpha\in[2.35,\, 1.35]$. These IMFs are all more top heavy than the fiducial IMF, but retain the same low-mass behavior $\alpha = 1.3$ for $M_\star< 0.5 M_\odot$, as this is significantly more well-constrained by observation than the high-mass behavior. 
In this (sub)section,  we assume a constant Star Formation Rate throughout the history of the galaxy; in Section 5.2 we will turn to other possibilities for the Star Formation History.

In the left panel of Fig.~\ref{fig:1}, we present a comparison of the values of $\epsilon$ between these top-heavy IMFs and the value of $\epsilon$ calculated using a fiducial Salpeter-like IMF. Through equation~\ref{eq:1}, this comparison is equivalent to a comparison between the inferred mass of a high redshift galaxy assuming a top-heavy IMF $M_{\rm lum}$ and the mass of the same galaxy inferred assuming a fiducial Salpeter-like IMF ($M_{\rm fid})$.
The vertical dotted and dash dotted lines mark the time since the formation of the galaxy to be $t=55\,$Myr and $t=102\,$Myr respectively, as these are canonical
times for studying the resulting star formation. The value of the corresponding redshift depends on the formation time of the galaxy.  As mentioned above, 
 if  the galaxy formed at $z=13.5$, these times correspond to redshifts $z\sim 12$ and $z\sim 11$ respectively; or if the galaxy formed at $z=17$, these vertical bars correspond to $z \sim 15$ and $z=13$; these bars therefore cover an interesting epoch relevant for JWST observations of distant galaxies.

We examine the evolution of the $\epsilon$ and $M_{\rm lum}$ factors over the course of the galaxy's life. We can immediately identify a few features of note. First, as predicted, the more top-heavy the IMF, generally the lower the factor $\epsilon/\epsilon_{\rm fid}$. In other words, if we assume a top-heavy IMF, a galaxy with the same overall mass and star formation rate will appear brighter. Equivalently, an observed galaxy with a fixed luminosity can be achieved with a lower star formation rate (for a given total halo mass). Furthermore, this relation appears to hold true at all times; there is no point where $\epsilon/\epsilon_{\rm fid}$ of a galaxy with a more top-heavy IMF is greater than that of a galaxy with a less top-heavy IMF at the same time.

What is perhaps the most notable feature of this evolution is the particular evolution that occurs for the most extreme top-heavy IMFs. One can see that $\epsilon/\epsilon_{\rm fid}$ reaches its minimum value at a galaxy age of approximately $t=300$~Myr  ($z\sim8.3$ for a galaxy that formed at z=13.5), where  the smallest value is $\epsilon/\epsilon_{\rm fid} \sim 1/15$ for the most top heavy IMF (largest  $\alpha$). Thus if we assume an IMF with a high-mass power law behavior of $dN_\star/dM_\star\sim M_\star^{-1.35}$, we could potentially lower the inferred star formation efficiency and equivalently $M_{\rm lum}$ by as much as a factor of 15, relative to 
the fiducial Salpeter high-mass IMF.  The large dip in the value of $\epsilon/\epsilon_{\rm fid}$ around this point, especially for the most top-heavy IMF galaxies can be explained as a result of two effects. First, the continuous formation of stars causes both the mass in stars $M_{\rm lum}$ and the luminosity of a galaxy $L_{\rm filter}$ to increase. While the most massive stars we consider begin to explode into supernova around $t\sim 5\,$Myr, the increasing abundance of lower mass, longer-lived stars causes the total galactic luminosity to increase for the first $\sim 300$~Myr. 
However, the luminosity scales as\footnote{The value of the exponent in this relation varies depending on the stellar mass, but is typically between about 3 and 6~\citep{kippenhahn}. The average value is around 3.5 for stars in the mass range $\sim 1-50~M_\odot$, and decreases to approximately 1 with increasing mass.} approximately $L_\star \sim M_\star^{3.5}$, so while the luminosity will continue to increase (and $\epsilon$ decrease) it will approach a constant value. 
However, at the same time, the explosions of high mass stars begins to enrich the metallicity in the interstellar medium. Therefore, later generations of stars will have higher initial metallicity than stars formed earlier. Generally, SSPs formed with a higher initial metallicity are redder~\cite{vanderbeke2013}, and so less of the luminosity they produce falls within the particular filter band we use. This effect causes $\epsilon$ to increase until the enhancements in a newly formed stars' initial metallicity no longer substantially shifts the spectrum, and the first generation of metal poor stars is removed from the galactic spectrum. 

Further, as one can see in Figure 1, the value of  $\epsilon/\epsilon_{\rm fid}$ is suppressed by roughly the same amount (within a factor of two)
regardless of the (high) redshift of the galaxies considered
(higher redshift corresponding to lower galactic age on the x-axis).  As previously mentioned, the largest suppression of 1/15 takes place for a galaxy at redshift 8.3 (for the case of galaxy formation at $z=13.5$)  As a comparative example, for the most top heavy $\alpha = 1.35$, JWST observations at $z=13$ (galactic age of 17 Myr for galaxy formation at $z=13.5$)  correspond to stellar galactic masses as low as 1/9 of what one would obtain assuming a Salpeter IMF.
Dropping the stellar mass of the galaxy by  $\sim 1/10$ helps to address the "too massive" problem of early JWST galaxies.

In addition to the evolution of $\epsilon$ over time, we also explore directly the relationship $\epsilon(\alpha)$ as shown in the right panel of Figure 1. While this relationship varies, particularly early in the evolution of the galaxy before star formation has reached an equilibrium, the relation between $\epsilon$ and $\alpha$ for $150\,{\rm Myr}\lesssim t\lesssim 250\,{\rm Myr}$ is largely the same, as shown by the black dashed line in the figure.\footnote{We also considered two larger ranges of galaxy ages ($150\,{\rm Myr}\lesssim t\lesssim 500\,{\rm Myr}$) and $150\,{\rm Myr}\lesssim t\lesssim 1000\,{\rm Myr}$ to see how these choices would change the fit. The variance around the best fit increases slightly, but the best fit exponential varies by only approximately 7\%.} 
A simple expression describes this relation and can be used to estimate how large of an impact changing the IMF could have on the inferred star formation efficiency. That is, if a galaxy is observed to have an inferred efficiency $\epsilon_{\rm fid}$ based on a Salpeter-like IMF, then a rough estimate for the star formation efficiency inferred when a more top-heavy IMF (subscript TH) is assumed would be 
\begin{equation}
    \epsilon_{\rm TH} = \epsilon_{\rm fid} e^{2.66(\alpha - 2.35)} \, ,
\end{equation}
or, equivalently, using Eq.~\eqref{eq:1}, 
\begin{equation}
    M_{{\rm lum}, \,{\rm TH}} = M_{{\rm lum}, \,{\rm fid} } e^{2.66(\alpha - 2.35)} \, ,
\end{equation}
where here the masses $M_{{\rm lum},\,{\rm TH}}$ and $M_{{\rm lum},\,{\rm fid} }$ refer to the masses inferred for a given galaxy using a top heavy IMF (with slope $\alpha$) and a Salpeter-like IMF, respectively. 
This relation is one of the most interesting results of this paper. 
For a galaxy with a given observed luminosity in JWST data, the required galactic stellar mass to explain the data decreases with more top heavy IMF slope $\alpha$ following this relation.
While SED fitting can in principle identify an IMF for an observed galaxy, in practice, the degeneracies introduced by allowing the IMF to vary make SED fitting challenging. As such, we suggest that the simple relationships given here are useful to identify an initial reasonable guess for the case of a top-heavy IMF, which can then be used in SED fitting for a more accurate parameter estimation.

\begin{figure*}
    \centering
    \includegraphics[width=\linewidth]{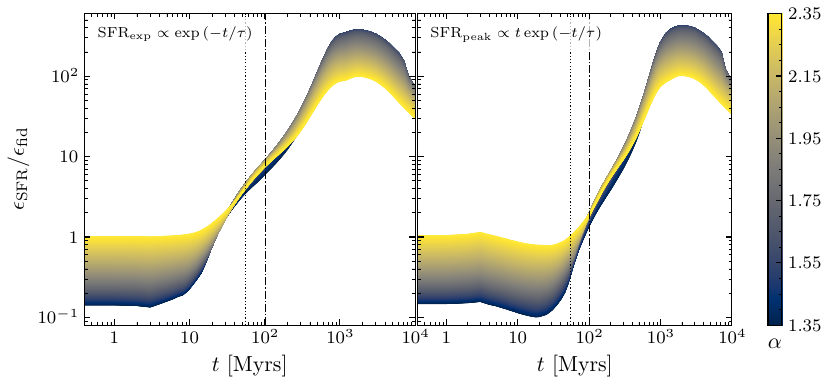}
    \caption{Effect of varying Star Formation Histories on Star Formation Efficiency (equivalently stellar galactic mass) inferred from JWST galaxy data. {\it Left}: Same as the left panel of Fig.~\ref{fig:1}, but for an exponentially decreasing star formation rate (SFR) with a characteristic timescale $\tau=100\,{\rm Myr}$ instead of a constant SFR. Here, all simulations assume an initial metallicity  $Z_{\rm ini}=0.001\,Z_{\odot}$. In the denominator $\epsilon_{\rm fid}$ corresponds to the fiducial Salpeter-like IMF ($\alpha=2.35$) and a constant SFR. As in the left panel of figure 1, different values of $\alpha$ in the IMF are represented by a distinct color as indicated by the colorbar to the right of the panel.
    {\it Right}: Same as the left panel, but for a peaked SFR with equivalent characteristic timescale $\tau=100\,{\rm Myr}$.  This early rise in the star formation rate briefly decreases the $\epsilon$ ratio before shifting to a phase resembling the exponential history in the left panel, with low-mass stars eventually dominating the luminosity. 
    Overall, one can see that decreasing SFR leads to larger SFE and $M_{\rm lum}$, competing with the effect of a top heavy IMF which goes in the opposite direction of lower $\epsilon$ and $M_{\rm lum}$, while increasing SFR (at low galactic ages in the right panel) does not counter the effect of a more top-heavy IMF.}
    \label{fig:2}
\end{figure*}

\subsection{Varying the Star Formation History}
So far the only change we have made to our fiducial model is the consideration of a more top-heavy IMF.
In this section we vary the star formation history (SFH), i.e. the Star Formation Rate as a function of time.
Up to this point we have only taken the star formation rate to remain constant over time. Now we consider two additional star formation histories, an exponential and a peaked history. 
As described previously, the ``exponential'' SFH  assumes an instantaneous onset to star formation followed by a star formation rate that decreases exponentially with time as ${\rm SFR}_{\rm exp}\propto \exp(-t/\tau)$ with time constant $\tau$ taken to be either 10 Myr or 100 Myr.  The ``peaked'' SFR is taken to initially rise with time (instead of an instaneous onset), followed by the same exponential decrease at late times.  In both cases, we normalize the star formation rate under the condition that the entire mass of the system is converted to stars over the course of 20,000 Myr.
The two time constants of $\tau = 10$~Myr, and $\tau=100$~Myr produced qualitatively similar results, but with the shorter timescale histories evolving more quickly.  Thus, as a representative example we plot only the 100~Myr timescale results in fig.~\ref{fig:2}. 

Fig.~\ref{fig:2} plots the ratio of the SFE efficiency for the exponential and peaked SFHs relative to the fiducial case of constant SFH as a function of the age of the galaxy.
We draw attention to three primary patterns that arise in the $\epsilon$ evolution over time. First, and perhaps most notably, whereas the $\epsilon$ ratio for top-heavy IMFs was always less than one in the constant-SFH case, when we allow for an exponential star formation history, the $\epsilon$ ratio can be substantially greater than 1. When star formation decreases, earlier generations of stars dominate the luminosity of a galaxy. Once massive stars are removed from the population, the remaining early stars, namely low-mass stars, will continue to dominate the stellar population. As a result, relative to a comparable constant SFH, galaxies with exponentially decreasing star formation histories tend to have a higher mass-to-light ratio and consequently higher $\epsilon$. However, what is also particularly noteworthy is that for an extensive period of time, roughly $20 - 800$~Myr after formation in the case of the exponential SFH with a time constant of $\tau=100$~Myr, the increase to $\epsilon$ due to this exponential SFH is effectively independent of the IMF chosen. Finally (at galactic ages of $t~1$~Gyr, corresponding to redshifts $z<5$, much lower than the focus of this paper), once the low-mass stars from the first generation begin to be removed from the population, the mass in stars decreases more rapidly than the luminosity from those stars, gradually shifting the mass-to-light ratio and $\epsilon$ downward. This leads to the notable slope at the late-time end of figure~\ref{fig:2}; the galactic age is so large at this point that it corresponds to $z<5$.
Overall, one can see that decreasing SFR leads to larger SFE and $M_{\rm lum}$, competing with the effect of a top heavy IMF which goes in the opposite direction of lower $\epsilon$ and $M_{\rm lum}$.

In the two panels of Figure~\ref{fig:2}, we compare the results for an ``exponential''  SFH with those of a ``peaked'' SFH. 
The two cases look very similar, as we would expect since the peaked history is defined to be the same as the exponential history beginning $100$~Myr after the formation of the galaxy. The most unique feature to the peaked results is a slight decrease in the ratio of $\epsilon$ approximately $3 - 30$~Myr after the formation of the galaxy. This suppression even affects the case involving the peaked SFH with the same IMF as the fiducial case, and so can be attributed to the SFH directly.  
Unlike the decreasing star formation rate during the exponential part of both the ``exponential'' and ``peaked'' cases, in the early part of the ``peaked'' SFR, the star formation rate increases with time. During that period, the increasing star formation rate is dominated by properties of the high-mass stars. Thus  
the top-heavy IMFs produce significantly lower $\epsilon$ ratios than the less top-heavy IMFs, as we also observed in the constant star formation case.

For galaxies that follow a peaked star formation history, the age of the galaxy plays an important role in its luminosity and consequently the inferred mass. Due to the top-heavy IMF, galaxies observed shortly after they were formed (within approximately one timescale $\tau$) would  have a star formation efficiency significantly lower than that predicted assuming a Salpeter IMF. Conversely, galaxies observed later in their evolution, when the star formation rate is decreasing, would instead have a much higher star formation efficiency than in the fiducial case. For example, we can consider two galaxies that formed at redshift $z=13.5$ with all the same properties except that we give one  a Salpeter-like IMF and a constant star formation history, whereas   the other  follows a peaked star formation history with timescale $\tau = 100$~Myr and has a top-heavy IMF with slope $\alpha=1.35$. If we were to observe these galaxies at redshift $z=12$, they would have an age of approximately $t=55$~Myr, during which the star formation in the galaxy following the peaked star formation history would be increasing, and the top-heavy galaxy would appear approximately 3 times as bright as the fiducial galaxy. If, however, these two galaxies were observed at redshift $z=11$ when they were each just over 100 Myr old, the top-heavy and fiducial galaxies would appear to have approximately the same brightness. And if observed at redshift $z=10$, when they were approximately 170 Myr old, the fiducial galaxy would appear approximately 3 times as bright as the top-heavy galaxy. 
As a consequence, the same galaxy (that formed at $z=13.5$)  would appear unusually faint at $z=10$ but unusually bright at $z=12$. We wonder if this effect could lead to a pile-up of galaxies observed to have unusually high luminosities at some high redshift; this would require a study of numbers of galaxies of different masses and ages at different redshifts, a study we leave to future work.

\begin{figure*}
    \centering
    \includegraphics[width=\linewidth]{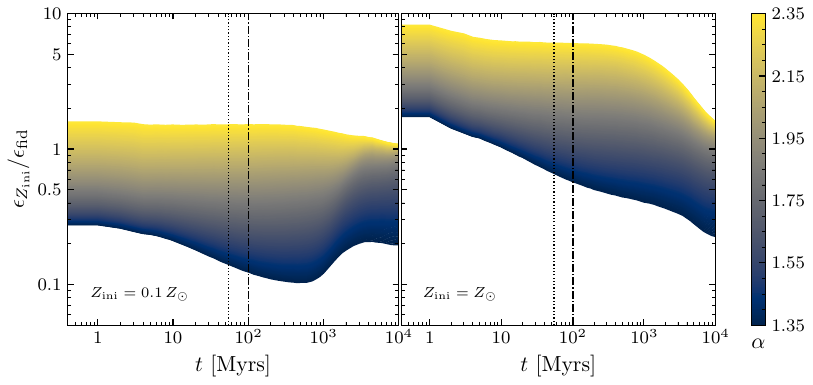}
    \caption{Same as the left panel of Fig.~\ref{fig:1}, but for higher initial metallicities of $Z_{\rm ini} = 0.1\,Z_\odot$ ({\it left}) and $Z_{\rm ini} =\,Z_\odot$ ({\it right}). Here, all simulations assume a constant star formation rate, with $\epsilon_{\rm fid}$ corresponding to the fiducial Salpeter-like IMF ($\alpha=-2.35$) and $Z_{\rm ini} = 0.001\,Z_\odot$. In both high-metallicity cases, the general trend of a more top-heavy IMF yielding lower values of $\epsilon/\epsilon_{\rm fid}$ persists, as in Figs. 1 and 2. However, the values of $\epsilon$ are higher for these high metallicity cases compared to the more  realistic lower-metallicity scenario, due to the redder spectra of high-metallicity stars, which emit less light in the JWST filter band used here.}
    \label{fig:3}
\end{figure*}

\subsection{Varying the Metallicity}
Finally, we consider the role that metallicity has on the $\epsilon$ and $M_{\rm lum}$ ratios. In all previous sections, we  used the fiducial value of $10^{-3}~Z_\odot$. Here we instead consider galaxies with constant star formation histories and unphysically large initial metallicities $Z_{\rm ini}$ of $0.1~Z_\odot$ and $1~Z_\odot$.  We plot the evolution of the ratios of $\epsilon/\epsilon_{\rm fid} = M_{\rm lum}/M_{\rm fid}$ as a function of the age of a galaxy $t$  in figure~\ref{fig:3}. 
In the case of  the Salpeter-like IMF, the ratios of $\epsilon$ for both high metallicity cases plotted here are greater than 1, and the ratio in the $Z_{\rm ini}=Z_\odot$ metallicity case is significantly higher than in the corresponding $Z_{\rm ini}=0.1~Z_\odot$. This supports the conclusion mentioned above that the increase in the ratio of $\epsilon$ seen at late times in figure~\ref{fig:1} can be attributed to the increasing metallicity of the interstellar medium. In addition, in both high-metallicity cases, the ratio of $\epsilon$ decreases as the galaxy evolves. This phenomenon can be explained as a result of higher-metallicity stars having longer lifetimes~\citep{bazan1990}. As the lifetimes of, in particular, massive stars increase, the luminosity of the galaxy will increase, and the ratio of $\epsilon$ will decrease. This has relatively little impact on the more realistic metallicity case, as they simply have not evolved long enough for high-metallicity stars to have a noticeable impact on the ratio of $\epsilon$, at least compared to the overall reddening of the galaxy that is also caused by increasing metallicity. Finally, even in these high-metallicity galaxies, we note the decrease in the ratio of $\epsilon$ due to having a more top-heavy IMF is comparable to the decrease seen in the more  realistic metallicity scenario, with a top heavy IMF with a power-law slope of $\alpha = 1.35$ reaching roughly $\sim\mathcal{O}(10 \%)$ of the Salpeter-like IMF result.

\section{Conclusions}
\label{sec:conc}

Observations of early galaxies by the James Webb Space Telescope revealed several surprising results. Early galaxies appear to be far
more massive and numerous than expected, which has been interpreted as resulting from expectedly high star formation efficiencies. While the inferred efficiencies from the observations are not physically impossible, under standard asssumptions, some of the highest efficiencies would require nearly all of the mass in these galaxies to form stars, which remains challenging to explain. In light of this, we explore the possibility that the star formation efficiencies inferred from JWST observations are overestimates, as a result of three possible effects: a top heavy Initial Mass Function (IMF) in lieu
of the standard Salpeter IMF; exponentially decreasing or peaked star formation rates in lieu of constant; and varying the initial metallicity. 

We used the \textsc{P\'egase} stellar population code to study the modified star formation efficiency, or equivalently inferred stellar galactic mass, that results from these effects. 
We find that allowing the high-mass end of the IMF to follow a power law with a greater (more top-heavy) slope than in the commonly used Salpeter IMF, the inferred star formation efficiency can be considerably lower. For example, if the high-mass end of the IMF goes as $M_\star^{-1.35}$, the star formation efficiency could be lower by a factor of $\sim 10$ than the star formation efficiency inferred under the assumption of a Salpeter-like power law of $M_\star^{-2.35}$. Furthermore, we considered whether different star formation histories or initial metallicities have an impact on this result. We found that in star formation histories where the star formation rate is constant or increasing over time, we see that top-heavy IMFs have similar reductions in their star formation efficiency. In case of star formation histories where the star formation rate is decreasing sufficiently quickly with time, 
for a standard IMF the inferred star formation rate is actually increased relative to the fiducial case, and even a top heavy IMF can no longer compensate for this increase.
Separately, the reduction in star formation efficiency due to a top-heavy IMF does not appear to be significantly impacted by changing the initial metallicity.

Finally, we leverage the obtained results to propose a simple  relationship between the reduction in star formation efficiency and the high-mass slope of the IMF. If the high-mass end of the IMF is described as $dN_\star/dM_\star \propto M_\star^{-\alpha}$, then the  modified star formation efficiency relative to the fiducial one, obtained using a Salpeter IMF for
constant star formation rate, can be parametrized as:
\begin{equation}
    \epsilon(\alpha) /\epsilon_{\rm fid} \approx  e^{2.66(\alpha - 2.35)},
\end{equation}
corresponding to an expression for the modified inferred galactic stellar mass relative to the one obtained using fiducial parameters,
\begin{equation}
    M_{\rm lum}(\alpha) /M_{{\rm lum}, \, {\rm fid}} \approx  e^{2.66(\alpha - 2.35)},
\end{equation}
This relation may be useful to estimate the impact that changing the IMF would have on observables.

Due to the various effects studied in this paper, galaxies of a given luminosity in JWST may be interpreted as having a larger stellar mass than they actually do.
Since there are more low mass galaxies than high mass galaxies, these effects may result in a larger number of seemingly overly massive galaxies than had been expected from the standard $\Lambda$CDM cosmology.
Thus, the effects studied in this paper may explain both puzzling observations for high luminosity galaxies in JWST:
the galaxies that have been interpreted as being  overly massive  as well as the larger than expected numbers of apparently high mass galaxies

\section*{Acknowledgements}

K.F.\ is Jeff \& Gail Kodosky Endowed Chair in Physics at the University of Texas at Austin, and K.F.\ , G.M.\ and J.J.Z.\ are grateful for support via this Chair. K.F., G.M., and J.J.Z.\  acknowledge support by the U.S.\ Department of Energy, Office of Science, Office of High Energy Physics program under Award Number DE-SC-0022021, as well as the Swedish Research Council (Contract No.~638-2013-8993).

%%%%%%%%%%%%%%%%%%%%%%%%%%%%%%%%%%%%%%%%%%%%%%%%%%
\section*{Data Availability}

The data underlying this article will be shared on reasonable request to the corresponding author.

\bibliographystyle{mnras}
\bibliography{jwst} 

\begin{thebibliography}{}
\makeatletter
\relax
\def\mn@urlcharsother{\let\do\@makeother \do\$\do\&\do\#\do\^\do\_\do\%\do\~}
\def\mn@doi{\begingroup\mn@urlcharsother \@ifnextchar [ {\mn@doi@} {\mn@doi@[]}}
\def\mn@doi@[#1]#2{\def\@tempa{#1}\ifx\@tempa\@empty \href {http://dx.doi.org/#2} {doi:#2}\else \href {http://dx.doi.org/#2} {#1}\fi \endgroup}
\def\mn@eprint#1#2{\mn@eprint@#1:#2::\@nil}
\def\mn@eprint@arXiv#1{\href {http://arxiv.org/abs/#1} {{\tt arXiv:#1}}}
\def\mn@eprint@dblp#1{\href {http://dblp.uni-trier.de/rec/bibtex/#1.xml} {dblp:#1}}
\def\mn@eprint@#1:#2:#3:#4\@nil{\def\@tempa {#1}\def\@tempb {#2}\def\@tempc {#3}\ifx \@tempc \@empty \let \@tempc \@tempb \let \@tempb \@tempa \fi \ifx \@tempb \@empty \def\@tempb {arXiv}\fi \@ifundefined {mn@eprint@\@tempb}{\@tempb:\@tempc}{\expandafter \expandafter \csname mn@eprint@\@tempb\endcsname \expandafter{\@tempc}}}

\bibitem[\protect\citeauthoryear{{Bazan} \& {Mathews}}{{Bazan} \& {Mathews}}{1990}]{bazan1990}
{Bazan} G.,  {Mathews} G.~J.,  1990, \mn@doi [\apj] {10.1086/168721}, \href {https://ui.adsabs.harvard.edu/abs/1990ApJ...354..644B} {354, 644}

\bibitem[\protect\citeauthoryear{Boylan-Kolchin}{Boylan-Kolchin}{2023}]{Boylan_Kolchin_2023}
Boylan-Kolchin M.,  2023, \mn@doi [Nature Astronomy] {10.1038/s41550-023-01937-7}, 7, 731–735

\bibitem[\protect\citeauthoryear{{Casey} et~al.,}{{Casey} et~al.}{2024}]{Casey2024}
{Casey} C.~M.,  et~al., 2024, \mn@doi [\apj] {10.3847/1538-4357/ad2075}, \href {https://ui.adsabs.harvard.edu/abs/2024ApJ...965...98C} {965, 98}

\bibitem[\protect\citeauthoryear{Chworowsky et~al.,}{Chworowsky et~al.}{2023}]{chworowsky2023}
Chworowsky K.,  et~al., 2023, Evidence for a Shallow Evolution in the Volume Densities of Massive Galaxies at $z=4$ to $8$ from CEERS (\mn@eprint {arXiv} {2311.14804}), \url {https://arxiv.org/abs/2311.14804}

\bibitem[\protect\citeauthoryear{Cueto, Hutter, Dayal, Gottlöber, Heintz, Mason, Trebitsch  \& Yepes}{Cueto et~al.}{2024}]{Cueto_2024}
Cueto E.~R.,  Hutter A.,  Dayal P.,  Gottlöber S.,  Heintz K.~E.,  Mason C.,  Trebitsch M.,   Yepes G.,  2024, \mn@doi [Astronomy &amp; Astrophysics] {10.1051/0004-6361/202349017}, 686, A138

\bibitem[\protect\citeauthoryear{Fioc \& Rocca-Volmerange}{Fioc \& Rocca-Volmerange}{2019}]{Fioc_2019}
Fioc M.,  Rocca-Volmerange B.,  2019, \mn@doi [Astronomy &amp; Astrophysics] {10.1051/0004-6361/201833556}, 623, A143

\bibitem[\protect\citeauthoryear{{Guo} et~al.,}{{Guo} et~al.}{2023}]{Guo2023}
{Guo} Y.,  et~al., 2023, \mn@doi [\apjl] {10.3847/2041-8213/acacfb}, \href {https://ui.adsabs.harvard.edu/abs/2023ApJ...945L..10G} {945, L10}

\bibitem[\protect\citeauthoryear{Kippenhahn, Weigert  \& Weiss}{Kippenhahn et~al.}{2012}]{kippenhahn}
Kippenhahn R.,  Weigert A.,   Weiss A.,  2012, Stellar Structure and Evolution, 2 edn.
Springer, \mn@doi{10.1007/978-3-642-3034-3}

\bibitem[\protect\citeauthoryear{{Labb{\'e}} et~al.,}{{Labb{\'e}} et~al.}{2023}]{Labbe2023}
{Labb{\'e}} I.,  et~al., 2023, \mn@doi [\nat] {10.1038/s41586-023-05786-2}, \href {https://ui.adsabs.harvard.edu/abs/2023Natur.616..266L} {616, 266}

\bibitem[\protect\citeauthoryear{Lapi et~al.,}{Lapi et~al.}{2024}]{lapi2024}
Lapi A.,  et~al., 2024, Constraining the Initial Mass function in the Epoch of Reionization from Astrophysical and Cosmological data (\mn@eprint {arXiv} {2403.07401}), \url {https://arxiv.org/abs/2403.07401}

\bibitem[\protect\citeauthoryear{Lu, Frenk, Bose, Lacey, Cole, Baugh  \& Helly}{Lu et~al.}{2024}]{lu2024}
Lu S.,  Frenk C.~S.,  Bose S.,  Lacey C.~G.,  Cole S.,  Baugh C.~M.,   Helly J.~C.,  2024, A comparison of pre-existing $\Lambda$CDM predictions with the abundance of {\it JWST} galaxies at high redshift (\mn@eprint {arXiv} {2406.02672}), \url {https://arxiv.org/abs/2406.02672}

\bibitem[\protect\citeauthoryear{McKinney et~al.,}{McKinney et~al.}{2024}]{mckinney2024}
McKinney J.,  et~al., 2024, SCUBADive I: JWST+ALMA Analysis of 289 sub-millimeter galaxies in COSMOS-Web (\mn@eprint {arXiv} {2408.08346}), \url {https://arxiv.org/abs/2408.08346}

\bibitem[\protect\citeauthoryear{Menon, Lancaster, Burkhart, Somerville, Dekel  \& Krumholz}{Menon et~al.}{2024}]{menon2024}
Menon S.~H.,  Lancaster L.,  Burkhart B.,  Somerville R.~S.,  Dekel A.,   Krumholz M.~R.,  2024, The Interplay between the IMF and Star Formation Efficiency through Radiative Feedback at High Stellar Surface Densities (\mn@eprint {arXiv} {2405.00813}), \url {https://arxiv.org/abs/2405.00813}

\bibitem[\protect\citeauthoryear{{Myers}, {Dame}, {Thaddeus}, {Cohen}, {Silverberg}, {Dwek}  \& {Hauser}}{{Myers} et~al.}{1986}]{myers1986}
{Myers} P.~C.,  {Dame} T.~M.,  {Thaddeus} P.,  {Cohen} R.~S.,  {Silverberg} R.~F.,  {Dwek} E.,   {Hauser} M.~G.,  1986, \mn@doi [\apj] {10.1086/163909}, \href {https://ui.adsabs.harvard.edu/abs/1986ApJ...301..398M} {301, 398}

\bibitem[\protect\citeauthoryear{{Offner}, {Clark}, {Hennebelle}, {Bastian}, {Bate}, {Hopkins}, {Moraux}  \& {Whitworth}}{{Offner} et~al.}{2014}]{offner2014}
{Offner} S.~S.~R.,  {Clark} P.~C.,  {Hennebelle} P.,  {Bastian} N.,  {Bate} M.~R.,  {Hopkins} P.~F.,  {Moraux} E.,   {Whitworth} A.~P.,  2014, in {Beuther} H.,  {Klessen} R.~S.,  {Dullemond} C.~P.,   {Henning} T.,  eds, Protostars and Planets VI. pp 53--75 (\mn@eprint {arXiv} {1312.5326}), \mn@doi{10.2458/azu_uapress_9780816531240-ch003}

\bibitem[\protect\citeauthoryear{Pacifici et~al.,}{Pacifici et~al.}{2023}]{Pacifici_2023}
Pacifici C.,  et~al., 2023, \mn@doi [The Astrophysical Journal] {10.3847/1538-4357/acacff}, 944, 141

\bibitem[\protect\citeauthoryear{{Pallottini} \& {Ferrara}}{{Pallottini} \& {Ferrara}}{2023}]{Pallottini2023}
{Pallottini} A.,  {Ferrara} A.,  2023, \mn@doi [\aap] {10.1051/0004-6361/202347384}, \href {https://ui.adsabs.harvard.edu/abs/2023A&A...677L...4P} {677, L4}

\bibitem[\protect\citeauthoryear{Peters, Schleicher, Smith, Schmidt  \& Klessen}{Peters et~al.}{2014}]{peters2014}
Peters T.,  Schleicher D. R.~G.,  Smith R.~J.,  Schmidt W.,   Klessen R.~S.,  2014, \mn@doi [Monthly Notices of the Royal Astronomical Society] {10.1093/mnras/stu1097}, 442, 3112

\bibitem[\protect\citeauthoryear{{Press} \& {Schechter}}{{Press} \& {Schechter}}{1974}]{1974press-schechter}
{Press} W.~H.,  {Schechter} P.,  1974, \mn@doi [\apj] {10.1086/152650}, \href {https://ui.adsabs.harvard.edu/abs/1974ApJ...187..425P} {187, 425}

\bibitem[\protect\citeauthoryear{{Rieke} et~al.,}{{Rieke} et~al.}{2008}]{Rieke2008}
{Rieke} G.~H.,  et~al., 2008, \mn@doi [\aj] {10.1088/0004-6256/135/6/2245}, \href {https://ui.adsabs.harvard.edu/abs/2008AJ....135.2245R} {135, 2245}

\bibitem[\protect\citeauthoryear{{Salpeter}}{{Salpeter}}{1955}]{Salpeter1955}
{Salpeter} E.~E.,  1955, \mn@doi [\apj] {10.1086/145971}, \href {https://ui.adsabs.harvard.edu/abs/1955ApJ...121..161S} {121, 161}

\bibitem[\protect\citeauthoryear{Shen, Vogelsberger, Boylan-Kolchin, Tacchella  \& Kannan}{Shen et~al.}{2023}]{shen2023}
Shen X.,  Vogelsberger M.,  Boylan-Kolchin M.,  Tacchella S.,   Kannan R.,  2023, The impact of UV variability on the abundance of bright galaxies at $z \geq 9$ (\mn@eprint {arXiv} {2305.05679}), \url {https://arxiv.org/abs/2305.05679}

\bibitem[\protect\citeauthoryear{Shuntov et~al.,}{Shuntov et~al.}{2025}]{Shuntov_2025}
Shuntov M.,  et~al., 2025, \mn@doi [Astronomy &amp; Astrophysics] {10.1051/0004-6361/202452570}, 695, A20

\bibitem[\protect\citeauthoryear{{Speagle}, {Steinhardt}, {Capak}  \& {Silverman}}{{Speagle} et~al.}{2014}]{Speagle2014}
{Speagle} J.~S.,  {Steinhardt} C.~L.,  {Capak} P.~L.,   {Silverman} J.~D.,  2014, \mn@doi [\apjs] {10.1088/0067-0049/214/2/15}, \href {https://ui.adsabs.harvard.edu/abs/2014ApJS..214...15S} {214, 15}

\bibitem[\protect\citeauthoryear{Steinhardt, Kokorev, Rusakov, Garcia  \& Sneppen}{Steinhardt et~al.}{2023}]{Steinhardt_2023}
Steinhardt C.~L.,  Kokorev V.,  Rusakov V.,  Garcia E.,   Sneppen A.,  2023, \mn@doi [The Astrophysical Journal Letters] {10.3847/2041-8213/acdef6}, 951, L40

\bibitem[\protect\citeauthoryear{Sun, Faucher-Giguère, Hayward, Shen, Wetzel  \& Cochrane}{Sun et~al.}{2023}]{sun2023}
Sun G.,  Faucher-Giguère C.-A.,  Hayward C.~C.,  Shen X.,  Wetzel A.,   Cochrane R.~K.,  2023, Bursty Star Formation Naturally Explains the Abundance of Bright Galaxies at Cosmic Dawn (\mn@eprint {arXiv} {2307.15305}), \url {https://arxiv.org/abs/2307.15305}

\bibitem[\protect\citeauthoryear{Trinca, Schneider, Valiante, Graziani, Zappacosta  \& Shankar}{Trinca et~al.}{2022}]{Trinca2022}
Trinca A.,  Schneider R.,  Valiante R.,  Graziani L.,  Zappacosta L.,   Shankar F.,  2022, \mn@doi [Monthly Notices of the Royal Astronomical Society] {10.1093/mnras/stac062}, 511, 616

\bibitem[\protect\citeauthoryear{Trinca, Schneider, Valiante, Graziani, Ferrotti, Omukai  \& Chon}{Trinca et~al.}{2024}]{trinca2024}
Trinca A.,  Schneider R.,  Valiante R.,  Graziani L.,  Ferrotti A.,  Omukai K.,   Chon S.,  2024, Exploring the nature of UV-bright $z \gtrsim 10$ galaxies detected by JWST: star formation, black hole accretion, or a non-universal IMF? (\mn@eprint {arXiv} {2305.04944}), \url {https://arxiv.org/abs/2305.04944}

\bibitem[\protect\citeauthoryear{Vanderbeke et~al.,}{Vanderbeke et~al.}{2013}]{vanderbeke2013}
Vanderbeke J.,  et~al., 2013, \mn@doi [Monthly Notices of the Royal Astronomical Society] {10.1093/mnras/stt2012}, 437, 1734

\bibitem[\protect\citeauthoryear{Woodrum et~al.,}{Woodrum et~al.}{2023}]{woodrum2023}
Woodrum C.,  et~al., 2023, JADES: Using NIRCam Photometry to Investigate the Dependence of Stellar Mass Inferences on the IMF in the Early Universe (\mn@eprint {arXiv} {2310.18464}), \url {https://arxiv.org/abs/2310.18464}

\bibitem[\protect\citeauthoryear{{Zubko}, {Dwek}  \& {Arendt}}{{Zubko} et~al.}{2004}]{Zubko_2004}
{Zubko} V.,  {Dwek} E.,   {Arendt} R.~G.,  2004, \mn@doi [\apjs] {10.1086/382351}, \href {https://ui.adsabs.harvard.edu/abs/2004ApJS..152..211Z} {152, 211}

\makeatother
\end{thebibliography}

\bsp	% typesetting comment
\label{lastpage}
\end{document}